\documentclass[aps,chaos,twocolumn,preprintnumbers,amsmath,amssymb]{revtex4}
\usepackage{graphicx,epsfig} % Include figure files
\usepackage{dcolumn} % Align table columns on decimal point
\usepackage{wrapfig}
\usepackage{bm} % bold math
\usepackage{color} % allow colours
\usepackage{ulem} % introduced by Sergio
\begin{document}

\title{Spatio-temporal dynamics induced by competing instabilities \\
 in two asymmetrically coupled nonlinear evolution equations}
\date{\today}

\author{D. Sch\"uler$^1$,  S. Alonso$^1$,  A. Torcini$^{2,3}$, and  M. B\"ar$^1$}

\affiliation {
$^1$ Physikalisch-Technische Bundesanstalt, Abbestrasse 2-12, 10587 Berlin, Germany \\
$^2$ CNR-Consiglio Nazionale delle Ricerche, Istituto dei Sistemi Complessi - 
Via Madonna del Piano 10, I-50019 Sesto Fiorentino, Italy\\
$^3$ INFN Sez. Firenze, via Sansone, 1 - I-50019 Sesto Fiorentino, Italy}

\begin{abstract}
Pattern formation often occurs in spatially extended physical, biological and chemical
systems due to an instability of the homogeneous steady state. The type of the
instability usually prescribes  the resulting spatio-temporal patterns and their
characteristic length scales. However, patterns resulting from the simultaneous occurrence 
of instabilities cannot be expected to be simple superposition of the
patterns associated with the considered instabilities. To address this issue we design two 
simple models composed by two asymmetrically coupled equations of non-conserved
(Swift-Hohenberg equations) or conserved  (Cahn-Hilliard equations) order
parameters with different characteristic wave lengths. The patterns arising in
these systems range from coexisting static patterns of different wavelengths to
traveling waves. A linear stability analysis allows to derive a two
parameter phase diagram for the studied models, in particular revealing for 
the Swift-Hohenberg equations a co-dimension two 
bifurcation point of Turing and wave instability and a region of coexistence of
stationary and traveling patterns. The nonlinear dynamics of the coupled evolution equations is
investigated by performing accurate numerical simulations. These reveal more complex patterns,
ranging from traveling waves with embedded Turing patterns domains to spatio-temporal chaos,
and a wide hysteretic region, where waves or Turing patterns coexist.
For the coupled Cahn-Hilliard equations the presence of an weak coupling is sufficient
to arrest the coarsening process and to lead to the emergence of purely periodic patterns.
The final states are characterized by domains with a
characteristic length, which diverges logarithmically with the coupling amplitude. 
\end{abstract}

\maketitle

%\textcolor{red}{ \bf 
{\bf Some chemical and biological systems exhibit competing pattern forming 
instabilities with different characteristic wave numbers. Often such a 
phenomenon is caused by the presence of different physical processes that appear 
on different length scales and cause patterns with different wavelengths. Here, 
we investigate two coupled Swift-Hohenberg (CH) equations  as well as two 
coupled Cahn-Hilliard (CH) equations as minimal models for such multiscale 
pattern formation. The CH and the SH equations 
are partial differential equations describing the evolution of
a conserved and a a non-conserved order parameter, respectively.
While the spatial domains in the SH equation self-organize into stationary periodic structures, for the CH equation 
the domains exhibit a coarsening dynamics that finally yield
a single large domain. The competition between two instabilities
with different wavelengths $\lambda_1$ and  $\lambda_2$ is analyzed for coupled 
SH equations as well as for coupled CH equations. In both cases, the coupling of 
equations with stationary instabilities (Turing or phase separation) can lead to 
wave dynamics. Moreover,  coupled SH equations exhibit a region of 
coexistence of Turing  and traveling patterns as well as more complex patterns.
The coupling of two CH equations leads to the arrest of coarsening and to the 
emergence of spatially periodic patterns.
}

\section{Introduction}

Reaction-diffusion equations are often employed to model systems outside of thermodynamic
equilibrium. In some cases, these systems self-organize to form
spatio-temporal structures \cite{cross1993}. 
Prominent examples of such phenomena are oscillatory chemical
reactions: e.g., the Belousov-Zhabotinsky (BZ) reaction produces
oscillations and waves in the concentration of the involved chemical species and
the CDIMA reaction produces stationary (Turing) patterns ~\cite{bookKapralShowalter}. 
Similar concepts of self-organization have been applied to explain different phenomena 
in biology~\cite{murray,keener}.

The linear stability analysis of such equations may reveal possible instabilities
in reaction-diffusion systems. Simple chemical and biochemical reactions can become unstable via
a Hopf bifurcation and produce oscillatory behaviour.
The addition of a spatial coordinate allows for the diffusion of the species and may
produce stationary periodic patterns via a Turing instability \cite{turing} or the emergence
of (traveling) waves via a wave instability~\cite{turing,hata}. Codimension-two bifurcations correspond to 
particular combinations of the parameter
values where  two types of instability appear simultaneously. In the proximity of such
points, the associated dynamics have been extensively studied in the case of Turing-Hopf 
codimension-two bifurcation~\cite{dewit1996,just2001,meixner1997,yang2004}, and analyzed
for the Turing-wave codimension-two bifurcation in a few instances~\cite{yang2002,alonso2011}. 
In particular, 
in the latter case, two instabilities appear simultaneously with two 
characteristic spatial scales, which can be very different depending on the 
parameter values. 
Such bifurcations appear in the BZ reaction \cite{yang2002}, in catalytic
surfaces with promotors \cite{john2013}, as well as in models of lipid domain
formation in biomembranes~\cite{john2005,john2005b}.
The interaction between different types of phospholipids and proteins on the
membrane of living cells induces a spatial instability of the
homogeneous state with a short characteristic spatial scale~\cite{john2005}.
On the other hand, the translocation of membrane proteins to the cytosol, where
they rapidly diffuse to a different location on the membrane, may cause a spatial
self-organization of the proteins on a larger spatial scale~\cite{john2005,alonso2010}. 
The coupling between these processes induce the emergence of oscillatory patterns at the membrane
at the larger spatial scale~\cite{john2005b}. This novel aspect is a central motivation in
setting up the models investigated in this paper.

\begin{figure}
\includegraphics[width=3.25in]{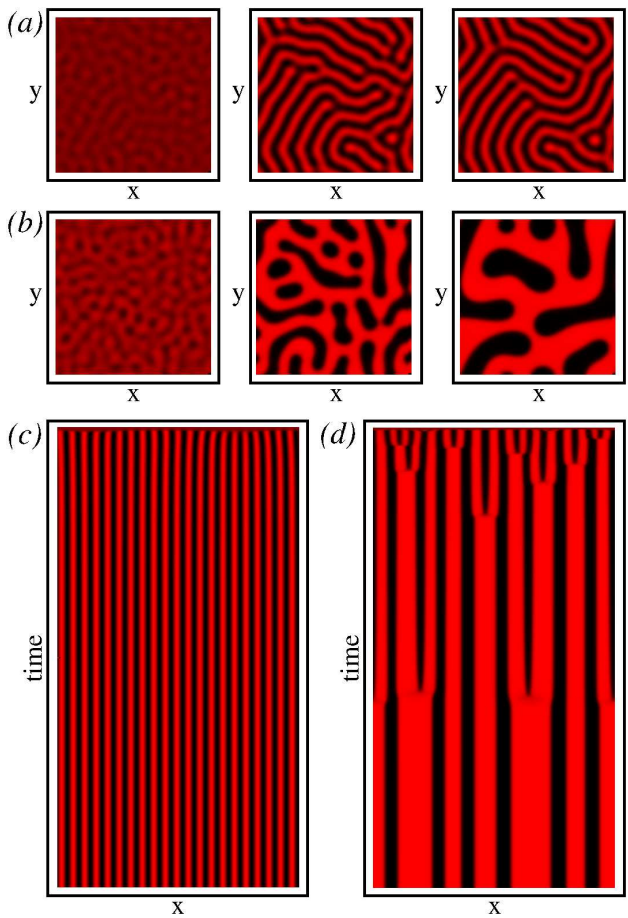}
\caption{Snapshots of a two-dimensional numerical simulations of a single 
SH equation (a) and a single CH equation (b) corresponding to  
times 8, 40 and 80. 
The total size of the system is $31 \times 31$.
Spatio-temporal plots of a one-dimensional numerical simulations of a single 
SH equation (c) and a single CH equation (d), total time 3000 and total size $L = 125$.
Parameter values are $k_1=2$ and $\varepsilon=0.4$. Time direction is going downwards in panels
(c) and (d).
}
\label{fig.0}
\end{figure}

The Swift-Hohenberg (SH) equation is a generic equation for a non-conserved order
parameter, which originally was developed for describing the instability of 
Rayleigh-Benard convection~\cite{swift1977,swift2008}. 
The SH equation was also applied to biology, for example as a model of nonlocal 
coupling in
biological systems describing neural tissues~\cite{hutt2005}. A complex variant 
of the SH equation has been previously employed to describe the 
dynamics of Class B lasers~\cite{lega1994}.
The standard SH equation undergoes a Turing instability following the increase of 
a bifurcation parameter, 
examples of the emerging patterns are reported in Figs.~\ref{fig.0}(a) and ~\ref{fig.0}(c) 
in one and two spatial dimensions. The structure of the SH equation permits 
a straightforward control the spatial scale of the instability by adjusting the
parameters entering in the equation.

The Cahn-Hilliard (CH) equation describes the process of phase separation in a
system with mass conservation~\cite{cahn1958}. The system
spontaneously segregates into spatial domains which grow and coarse continuously, see
Fig.~\ref{fig.0}(b) and ~\ref{fig.0}(d) for 2D and 1D examples.
The difference between the evolution of the single SH and CH equations with similar
characteristics can be appreciated in Fig.~\ref{fig.0}.

Here we employ two coupled SH (CH) equations to introduce two different spatial scales
in a system with two non-conserved (conserved) order parameters. 
The two equations are connected by an asymmetric coupling which induces a repertoire of spatio-temporal evolutions. 
While a model with a symmetric coupling between the two equations could be derived from an energy functional 
using a variational approach, asymmetrically coupled systems represent an effective description.
However, such asymmetric coupling permits us to generate simple and generic
models where two spatial bifurcations compete and give rise to waves.
The combination of several CH may describe the process of phase
separation with three or more components. In particular, it could be of
interest for the processes of lipid separation at membranes~\cite{veatch2003}
or of phase separation in block copolymers melt~\cite{Oono}.
The combination of two SH may mimic the coupling of two layers where Turing patterns appear \cite{yang2006,li2014}.

The present article is organized as follows.
In Sect. II the studied models are introduced and the corresponding linear
stability analysis reported. Sect. III is devoted to the presentation of the results 
obtained in this study. In particular, the linear stability diagrams for the two models
are described in Subsections III A, while subsection III C reports 
detailed numerical investigations of the two models. 
Finally, the main results of this study are summarized in Sect. IV.

\section{Models}

Two simple models of multiscale pattern formation, based on the asymmetric coupling 
of two identical SH resp. CH equations,  are the main subjects of this article and 
are introduced in this Section.

\begin{figure}
\includegraphics[width=3.25in]{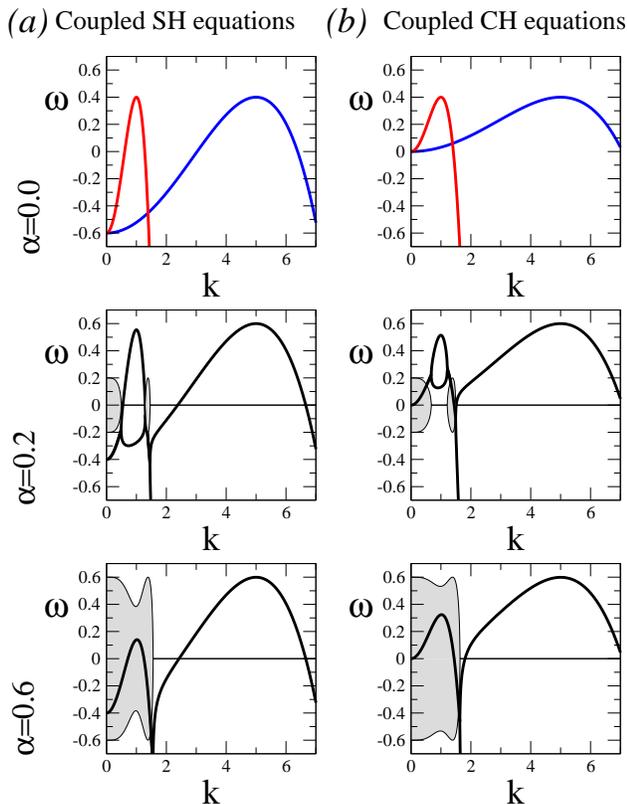}
\caption{Dispersion relation for two uncoupled (top) with $\alpha=0$ and coupled 
(middle) with $\alpha=0.2$ (bottom) with $\alpha=0.6$ SH (a) and CH (b) 
equations. Thick (thin) solid lines refer to real (imaginary) part of the eigenvalue 
$\omega$. Parameter values are $k_1=1$ and $k_2=5$ and $\varepsilon=0.6$. 
}
\label{fig.1}
\end{figure}

\begin{figure*}
\includegraphics[width=6.0in]{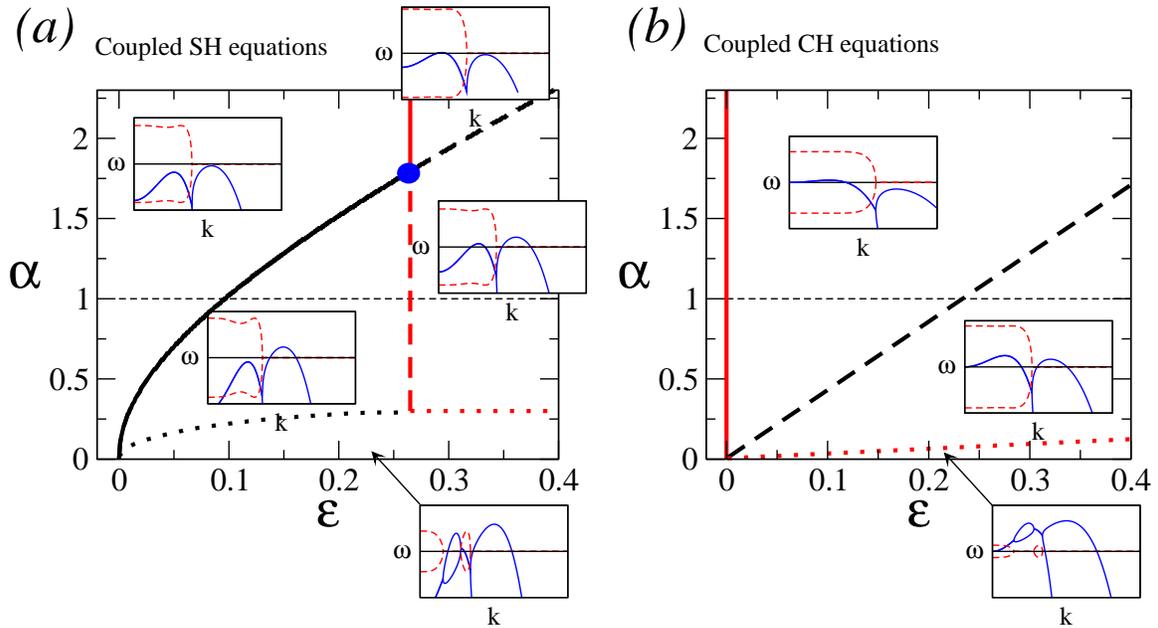}
\caption{Linear stability diagrams as given by Eqs. (\ref{eq.SH.2}a) and (\ref{eq.CH.2}b) for the parameters $\varepsilon$ and $\alpha$. 
Solid black (red) lines in (a) show Turing (wave) bifurcation,  whereas  the solid black lines 
in (b) correspond to emergence of an unstable band with finite wavenumber.
The black (red) dashed lines indicate the emergence of a secondary Turing (wave) instability. 
Dotted lines mark the transition from real to complex maximum in the dispersion relation, see Fig.~\ref{fig.1}.
Thick point indicates codimension-two point $(\varepsilon_c;\alpha_c)$. 
Insets show characteristics dispersion relations in the corresponding region, where solid (dashed) lines correspond 
to real (imaginary) eigenvalues. Thin dashed horizontal lines for comparison with Fig.~\ref{fig.4}.
Rest of parameter values are $k_1=1$ and $k_2=2$.}
\label{fig.2}
\end{figure*}

\subsection{Coupled SH Equations}

We start considering the single SH equation, that was originally derived from the equations 
for thermal convection, but is commonly used as a generic model of pattern formation \cite{swift2008,kapralbook}. 
The SH equation exhibits static Turing patterns similar to the one observed for reaction-diffusion equations 
of activator-inhibitor type. In particular, the SH equation describes the spatial evolution 
of a single non-conserved dynamic variable $u$: 
\begin{equation}
\label{eq.SH}
\frac{\partial u}{\partial t} = \varepsilon u - \left( \frac{1}{k^2_1} \nabla^2 +1 \right)^2  u - u^3,
\end{equation}
where the parameter $\varepsilon$ controls the linear stability of the homogeneous
stationary solution $u_0=0$ and the parameter $k_1$ is the critical characteristic wavenumber 
at the onset of instability at $ \varepsilon = 0 $. One can easily determine
the stability of the solution $u_0$ by the introduction of 
an infinitesimal spatially periodic perturbation, namely by considering $u = u_0 + \delta u \enskip e^{\omega t - i k x}$. 
The resulting dispersion relation$\omega(k)$ is real-valued and it depends on the wavenumber $k$ as follows
\begin{equation}
\label{eq.SH.dr}
\omega(k) = \varepsilon - 1 + 2  \frac{k^2}{k^2_1} - \frac{k^4}{k^4_1} \enskip .
\end{equation}
When $\varepsilon > 0$, the homogeneous state become unstable for a finite interval
of wavenumbers around $k_1$, namely for  
$\sqrt{1-\sqrt{\varepsilon}} \le k/k_1 \le \sqrt{1+\sqrt{\varepsilon}} $.
The expected characteristic spatial scale of the resulting Turing pattern is given by 
$\lambda_1 = 2 \pi / k_1$.

Next, we consider two linearly coupled SH equations, as simple model for multiscale pattern formation
\begin{eqnarray}
\frac{\partial u}{\partial t} =  \varepsilon u - \left( \frac{1}{k^2_1} \nabla^2 +1 \right)^2  u - u^3 - \alpha v \enskip, \nonumber \\
\frac{\partial v}{\partial t} =  \varepsilon v - \left( \frac{1}{k^2_2} \nabla^2 +1 \right)^2  v - v^3 + \alpha u \enskip .
\label{eq.SH.2}
\end{eqnarray}
Each of these equations has a different characteristic length
$\lambda_i = 2 \pi/k_i$, with $i =1,2$. Furthermore,  the same
control parameter $\varepsilon$  for the instability is used in both equations ensuring that the
instability occurs simultaneously in the decoupled systems. 
The coupling parameter $\alpha$ is the same in both equations. Note,  however the
opposite signs, which renders Eq. ~\eqref{eq.SH.2} non-variational, i. e. the dynamics of $u$ and $v$ 
in Eq.~\eqref{eq.SH.2} cannot be derived as variational derivates of some functional ${\cal F} (u,v)$. 
We analyze the stability of the homogeneous solution $u_0=v_0=0$ 
by considering the following perturbations $u=u_0+ \delta u \enskip e^{\omega t - i k x}$ and
$v=v_0+ \delta v e^{\omega t - i k x}$. The linear stability analysis leads to the following 
dispersion relation:
\begin{eqnarray}
\label{eq.SH.2.dr}
\omega(k) &=& \varepsilon - 1 + k^2 \frac{ k^2_1 + k^2_2}{k^2_1 k^2_2} -k^4 \frac{k^4_1 +  k^4_2}{2 k^4_1 k^4_2}  \\ 
&\pm & \sqrt{\left[k^2 \frac{k^2_2 - k^2_1}{k^2_1 k^2_2} - k^4 \frac{k^4_2 -  k^4_1}{2 k^4_1 k^4_2}\right]^2 -\alpha^2}
\enskip ,
\nonumber
\end{eqnarray}
which can produce spatial and spatio-temporal instabilities depending on the parameter values.
For $\alpha=0$, the two equations become uncoupled and two real dispersion
relations with maxima in $k_1$ and $k_2$ are obtained (see top panel of Fig.\ref{fig.1}(a)). 
By coupling the two systems the dispersion relation is modified.
Nevertheless the real part of the leading eigenvalue still resembles the respective curve for the uncoupled
system with $\alpha = 0$ and exhibits two maxima. The wavenumbers associated to these maxima will
be the dominant modes of the coupled dynamics. We will indicate them as mode 1 and mode 2, corresponding
to small and large wavenumbers, respectively. Furthermore, for sufficiently large
coupling, the leading eigenvalue becomes complex for low wavenumbers, see middle and bottom panel of Fig.\ref{fig.1}(a).

\subsection{Coupled CH Equations}

The standard CH equation is commonly used as a paradigmatic
model of phase separation~\cite{kapralbook}. In contrast to
SH equation, it describes the evolution of a single conserved variable $u$: 
\begin{equation}
\label{eq.CH}
\frac{\partial u}{\partial t} = \nabla^2 \left( u^3 - 2 \varepsilon  
\frac{1}{k^2_1}  u - \varepsilon  \frac{1}{k^4_1} \nabla^2 u \right) .
\end{equation}
Again, the spatially homogeneous state $u_0=0$ is a stationary solution of the equation.
By applying the perturbation $ \delta u \enskip e^{\omega t - i k x}$ and by
linearizing Eq. (\ref{eq.CH}) around $u_0$, one obtains the dispersion relation as:
\begin{equation}
\label{eq.CH.dr}
\omega(k) = 2 \varepsilon \frac{k^2}{k^2_1} - \varepsilon \frac{k^4}{k^4_1} \enskip .
\end{equation}
For $\varepsilon > 0$ the homogeneous state become unstable for a finite interval
of wavenumbers bounded from above from $k_1$,  namely for  $0 < k < \sqrt{2} k_1$.
The characteristic spatial scale of the initial pattern is given by the most unstable
wavenumber, namely $\lambda_1 = 2 \pi / k_1$. 
For $\varepsilon < 0$ the system is unstable for small spatial scales and higher order 
spatial derivatives are needed to stabilize the system.

Analogously to the previous analysis, we linearly couple two CH equations in an asymmetric way, as follows:
\begin{eqnarray}
\frac{\partial u}{\partial t} =  \nabla^2 \left( u^3 - 2 \varepsilon  \frac{1}{k^2_1}  u - \varepsilon  \frac{1}{k^4_1} \nabla^2 u \right) - \alpha v, \nonumber \\
\frac{\partial v}{\partial t} =  \nabla^2 \left( v^3 - 2 \varepsilon  \frac{1}{k^2_2}  v - \varepsilon  \frac{1}{k^4_2} \nabla^2 v \right) + \alpha u, 
\label{eq.CH.2}
\end{eqnarray}
as for the coupled SH equations, we assume the same value of the control parameter $\varepsilon$
for both equations and the same coupling $\alpha$ with opposite signs. Once more
we analyze the stability of the homogeneous solution $(u_0,v_0)=(0,0)$ by considering 
periodic perturbations to the vector $(u,v)$, namely $(\delta u \enskip e^{\omega t - i k x}, \delta v \enskip e^{\omega t - i k x})$.
The linear stability analysis leads to the following dispersion relation:
\begin{eqnarray}
\label{eq.CH.2.dr}
\omega(k) &=& \varepsilon k^2 \frac{ k^2_1 + k^2_2}{k^2_1 k^2_2} - \varepsilon k^4 \frac{k^4_1 +  k^4_2}{2 k^4_1 k^4_2}  \\ 
&\pm & \sqrt{\varepsilon^2 \left[  k^2 \frac{k^2_2 - k^2_1}{k^2_1 k^2_2} -  k^4 \frac{k^4_2 -  k^4_1}{2 k^4_1 k^4_2}\right]^2 -\alpha^2}
\enskip .
\nonumber
\end{eqnarray}
As shown in Fig.~\ref{fig.1}(b), also in this case the coupled system exhibits 
two real maxima in the dispersion relation, resembling that of the uncoupled system.
Furthermore, also in the present case we will denote the wavenumbers associated to these maxima as
mode 1 at small $k$ and mode 2 at large $k$.
The presence of the coupling between the two equations
induces the emergence of imaginary components in the dispersion relation 
in the low wavector part of the spectrum, as shown Fig.~\ref{fig.1}(b).

\section{Results}

In order to characterize the two previously introduced models,
we first analyze their linear stability diagrams. 
In addition, we have performed extensive numerical simulations of the full nonlinear models.

\subsection{Linear stability diagrams}
 
The linear stability analysis indicates that
different types of behaviours are expected for different choices of the 
parameters $\varepsilon$ and $\alpha$.
The linear stability diagrams for both systems are shown in Fig.~\ref{fig.2}.
As a general remark, for small value of the coupling $\alpha$ both systems present 
spatial patterns. However, these are atypical spatial patterns, 
due to the coexistence of real mode 1 and mode 2 with unstable complex modes 
associated to non critical wavenumbers, see panels for $\alpha=0.2$ in 
Fig.~\ref{fig.1}.
 
The linear stability diagram for the coupled SH equations is shown in Fig.~\ref{fig.2}(a). 
For  a large coupling constant $\alpha$ and small values of the control parameter $\varepsilon$, 
the homogeneous steady state is 
stable. This state can lose stability in two different ways, depending  $\alpha$ 
is larger or smaller than a critical value $\alpha_c$. For $\alpha > \alpha_c$ 
one observes a wave bifurcation involving mode 1 at
\begin{equation}
\label{epsC}
\varepsilon_c = \frac{1}{2} - \frac{k_1^2 k_2^2}{k_1^4 + k_2^4} \enskip ;
\end{equation}
that  corresponds to the solid red vertical line in Fig.~\ref{fig.2}(a). For $\alpha < \alpha_c$,
the system undergoes a Turing instability for mode 2 (solid black line in Fig.~\ref{fig.2}(a))
and at $\varepsilon=\varepsilon_c$ an additional band of unstable oscillatory modes around 
mode 1 emerges (red dashed line in Fig.~\ref{fig.2}(a)).
The two lines in the phase diagram, associated to these transitions, cross in a 
codimension-two point $(\varepsilon_c,\alpha_c)$ indicated by the blue dot in Fig.~\ref{fig.2}(a).
For decreasing $\alpha$-values,  at the dotted lines  Fig.~\ref{fig.2}(a), 
the eigenvalue associated to mode 1 passes from complex to real positive values, although other modes 
can still remain complex. 

The linear stability diagram for the coupled CH equations is shown in Fig.~\ref{fig.2} (b).
For positive $\varepsilon > 0$ the system exhibits a wave bifurcation (red solid vertical line in 
Fig.~\ref{fig.2}(b)). Furthermore, at sufficiently large $\alpha$ values mode 1 is unstable and complex. 
By decreasing the $\alpha$ parameter the system undergoes a secondary instability of mode 2 
 connected with real eigenvalues (black solid line in Fig.~\ref{fig.2}(b)).
Below this black line, these two types of unstable modes coexist. At sufficiently
small $\alpha$ values mode 1 becomes purely real (dotted red line in Fig.~\ref{fig.2}(b)),
although other modes are still complex.

\begin{figure}
\includegraphics[width=3.25in]{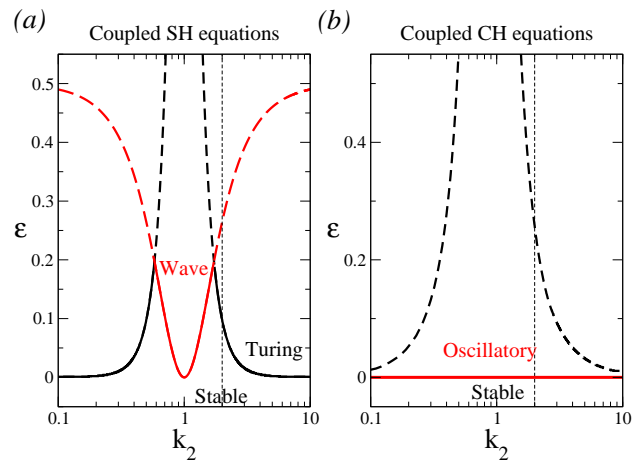}
\caption{Linear stability diagram as given by Eqs. (\ref{eq.SH.2}a) and (\ref{eq.CH.2}b)
for the parameters $\varepsilon$ and $k_2$, keeping 
$\alpha=1$ and $k_1=1$, for two coupled SH equations (a), and two coupled CH equations (b).
The lines have the same meaning as in Fig.~\ref{fig.3}.
Thin dashed vertical lines refer to the parameter $k_2$ employed in Fig. \ref{fig.2}.
}
\label{fig.4}
\end{figure}

In Fig.\ref{fig.4} we systematically change the ratio between the two parameters
$k_1$ and  $k_2$ for the two models. For the special case $k_1=k_2$, the traveling solution
is always present independently of the value of the other parameters (since the imaginary
components of $\omega (k)$ in Eqs. (\ref{eq.SH.2.dr}) and (\ref{eq.CH.2.dr}) is in this case 
constant and independent of $k$). The linear stability diagrams are symmetric around the point $k_1=k_2$.

For the coupled SH equations, the appearance of traveling solutions does not depend 
on the coupling strength but on the control parameter 
$\varepsilon$ (red lines in Fig.~\ref{fig.4}(a)). The critical value of the $\varepsilon$ parameter, 
controlling the emergence of waves, depends 
on the relation between the two scales $k_1$ and  $k_2$, and it can be analytically calculated
from Eq.(\ref{epsC}).
For $k_2>>k_1$ the secondary instability  occurs at $\varepsilon_c=0.5$, see red dashed line in Fig.~\ref{fig.4}(a). 
In between $0<\varepsilon_c < 0.5$ the dynamics depends on the coupling and $k_2$.

\begin{figure}
\includegraphics[width=3.25in]{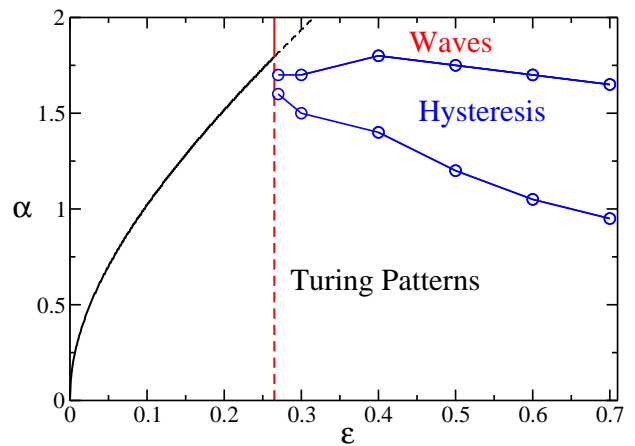}
\caption{Phase diagram $(\varepsilon;\alpha)$ for the coupled SH Eqs.(\ref{eq.SH.2})
estimated numerically. For the details on the numerical simulation see the text.
The solid and dashed black (red) lines refer to the results of the linear stability
analysis and have the same meaning as in Fig.~\ref{fig.3}. The blue line with symbols
denote the limits if the hysteretic region.
Parameter values are $k_1=1$ and $k_2=2$, the system has been integrated for each couple
of $(\varepsilon,\alpha)$ parameters for a time $20,000$ with a time step $\Delta t = 0.001$
by considering a system size $L = 40 \pi$ and by employing 512 Fourier modes.
}
\label{fig.6}
\end{figure}

The linear stability diagram for the coupled CH equations is a simplified version of the
previous case, as one can appreciate by comparing panels (a) and (b) in Fig.~\ref{fig.4}. 
The wave instability is always present, at least for not too large values of
$\varepsilon$, however, large $k_2$ enhances the appearance of the spatial
instability and promotes the competition between both instabilities.

\begin{figure}[]
\includegraphics[width=3.in]{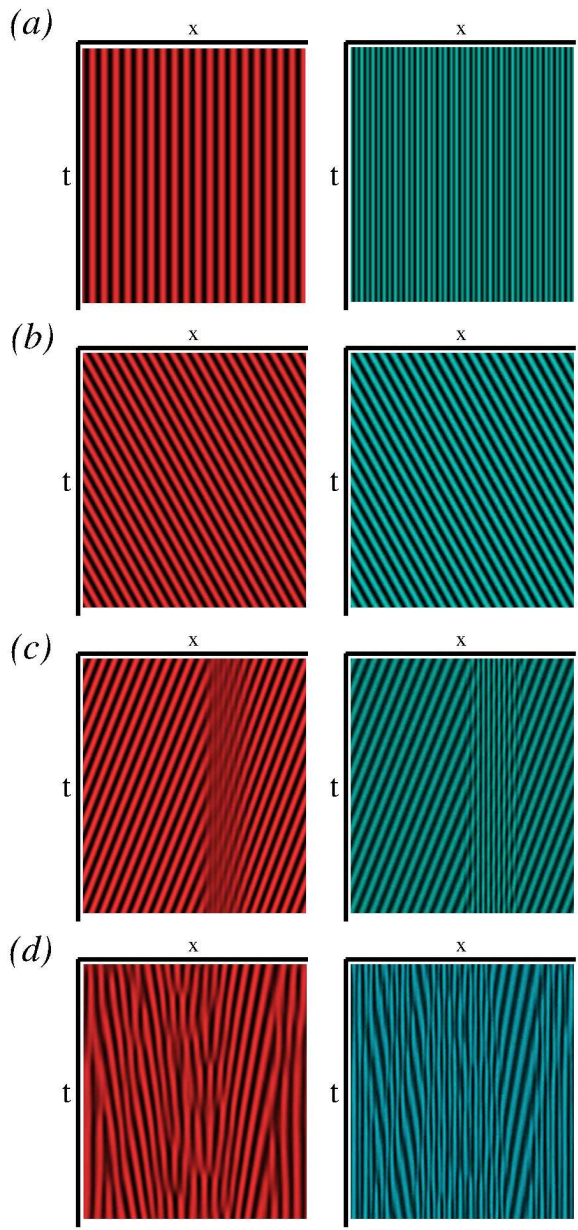}
\caption{Spatio-temporal plots of the dynamics of u (left) and v (right) variables obtained after numerical integration of 
the coupled SH Eqs. (\ref{eq.SH.2}) for the parameters $\varepsilon=0.3$ and 
$\alpha=0.3$ (a), $\varepsilon=0.4$ and $\alpha=1.6$ (b), $\varepsilon=0.7$ and 
$\alpha=1.2$ (c), and $\varepsilon=0.9$ and $\alpha=0.6$ (d). 
Rest of parameter values are $k_1=1$ and $k_2=2$. 
The integration time is $50$, after discarding a transient of $2000$ with $\Delta t = 0.001$, 
and the system size is set to $L = 40 \pi$ with a spatial discretization $\Delta x= L /512 \simeq 0.245$.
Time direction is going downwards in all panels.}
\label{fig.3}
\end{figure}

\subsection{Numerical simulations}

We have employed a time splitting pseudo spectral method, similar to the one described in \cite{torc_pol},
to numerically integrate Eqs. (\ref{eq.SH.2}) and (\ref{eq.CH.2}). The simulations have been 
performed by considering mainly one dimensional systems with size $L = 40 \pi$ or $L = 80 \pi$
with periodic boundary conditions.
For the numerical integration 512 or 1024 Fourier modes have been alternatively
used and integration time steps in the range
$\Delta t = 0.001 - 0.01$. The runs are usually initialized by 
setting $u$ and $v$ to random  values uniformly distributed 
in the interval $[-1;1]$.

The numerical simulations reproduce the results predicted by the linear
stability analysis for small values of $\alpha$, while for larger coupling $\alpha$  
nonlinear effects come into play leading to a richer scenario
not predicted by linear analysis. Here, we focus on the competition between  
waves and Turing patterns revealed by the coupled SH equations and on 
the arrest of coarsening occurring in the coupled CH equations.

\subsubsection{Coupled SH Equations}

The numerically obtained phase diagram for the coupled SH equations 
is shown in Fig.~\ref{fig.6}, which has been obtained by keeping constant 
$\varepsilon$ ($\alpha$) (for a certain set of values) and by varying the
other parameter, namely $\alpha$ ($\varepsilon$). The parameter $\alpha$ ($\varepsilon$) 
is first increased and successively decreased of a constant amount
$\Delta \alpha = 0.05$ ($\Delta \varepsilon = 0.05$).
Each simulation had a duration of $2,000 - 20,000$ time units and the next simulation 
in the sequence is initialized by employing the final state of the previous one.
This allows to reveal a hysteretic transition in the $(\varepsilon,\alpha)$-plane
for $\varepsilon > \varepsilon_c$: the corresponding hysteretic region is enclosed by the 
blue curves in Fig.~\ref{fig.6}. Therefore, within this region reported traveling waves 
or Turing patterns can be observed, depending on the initial conditions.

\begin{figure}
\includegraphics[width=3.in]{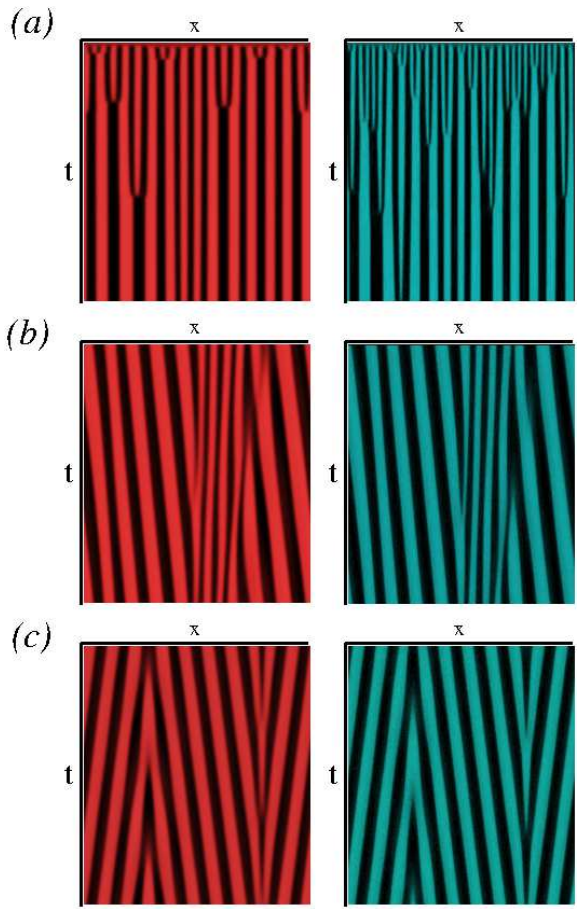}
\caption{Spatio-temporal plots of the dynamics of the $u$ (left) and $v$ (right) variables 
obtained after numerical integration of the coupled CH Eqs. (\ref{eq.CH.2}) for the parameters
$\varepsilon=0.5$ and $\alpha=0.002$ (a), $\alpha=0.15$ (b), and $\alpha=0.2$ (c).
The integration time is $1,500$ without any transient for (a) and $50$ after a 
transient of $2,000$ for (b) and (c) with $\Delta t = 0.001$, 
and system size is $L = 40 \pi$ and spatial discretization $\Delta x = L 
/512 = \simeq 0.245$. Time direction is going downwards in all panels.}
\label{fig.3b}
\end{figure}

Some examples of Turing patterns and waves obtained in the simulations are reported
in Fig.~\ref{fig.3}. When both systems are weakly coupled, i.e. small $\alpha$ in Fig.~\ref{fig.6},
two distinct Turing patterns characterized by different spatial scales can be observed 
in the variable $u$ and $v$ (Fig.~\ref{fig.3}(a)). 
%At this small coupling the larger spatial scale dominates and 
%Turing patterns appear in both subsystems with small wave-lengths. \red{IS IT TRUE ???}.
For larger $\alpha$, the coupling eliminates the instability with smaller
characteristic scale and generates a wave instability associated to the 
larger spatial scale in both variables, see Fig.\ref{fig.3}(b).
Deep inside the hysteretic region one can observe the coexistence of Turing
patterns, embedded in traveling waves (as shown in Fig.\ref{fig.3}(c)).
For very large $\varepsilon \simeq 0.9$ spatio-temporal chaotic dynamics
with defects is observable (see Fig.~\ref{fig.3}(d)) and we have verified that this is not a transient
regime by performing long simulations up to time $t \simeq 50,000$.

\subsubsection{Coupled CH Equations}

\begin{figure}
\includegraphics[width=3.in]{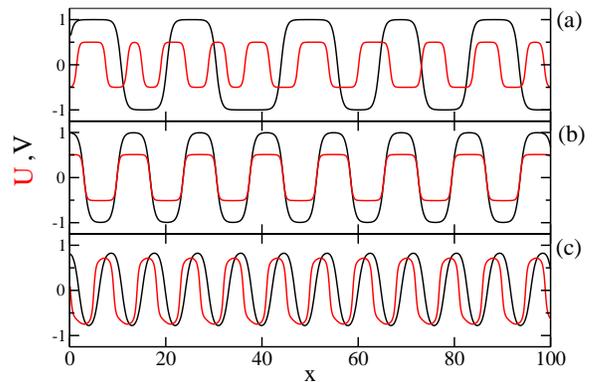}
\caption{Characteristic patterns for the $u$, $v$ variables in one spatial
dimension obtained by numerical integration of 
Eqs. (\ref{eq.CH.2}) for parameters $\varepsilon=0.5$ and 
$\alpha=0$ (a), $\alpha=0.002$ (b), $\alpha=0.2$ (c).
The systems size is  $L=80 \pi$ and the integration is performed by
employing a time splitting pseudo-spectral code with 
$1,024$ Fourier modes and an integration time step $\Delta t = 0.01$.
The configurations in (a) refer to an integration time $T=41,942$,
while those in (b) and (c) to a time $t=671,088$ 
}
\label{fig.9}
\end{figure}

Examples of patterns found in simulations of the coupled CH
equations, 
for increasing coupling parameter $\alpha$ are shown in Fig.~\ref{fig.3b}.
For small $\alpha$ one observes initial coarsening in the $u$ and $v$ variables similar to what is found in the uncoupled single SH equations Fig.\ref{fig.3b}(a). 
As expected from the linear stability analysis above, the initial domain patterns of $u$ and $v$ have different characteristic wavelength. 
However, the coarsening process stops after a finite time and both variables - $u$ and $v$ - exhibit a domain pattern of the same wavelength. 
For larger values of $\alpha$, are observed in line with the occurrence of oscillatory unstable modes in the linear stability analysis. 

traveling domains are observed as expected from the linear stability analysis, see Fig.~\ref{fig.3b}(b) and (c). 
Figure ~\ref{fig.9} shows typical $u$ and $v$ for the different cases. 
Independent domain patterns in the uncoupled equations in Fig.~\ref{fig.9}(a), patterns with equal wavelength in $u$ and $v$ for the case of arrested coarsening in Fig.~\ref{fig.9}(b) form small $\alpha$ and, finally, left-traveling domains in Fig.~\ref{fig.9}(c) for large $\alpha$.

\begin{figure}
\includegraphics[width=3.in]{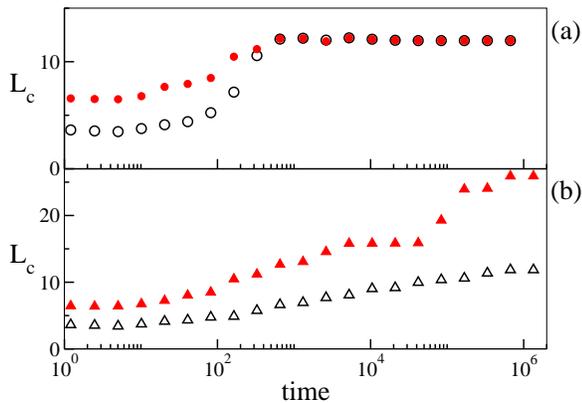}
\caption{Characteristic size $L_c$ of the domains obtained by numerical integration of 
Eqs. (\ref{eq.CH.2}) for the parameters $\varepsilon=0.5$ and $\alpha=0.002$ (a), $\alpha=0.0$ (b).
The (red) filled symbols refer to variable $u$, and the (black) empty to $v$.
The system size and integration details are as in Fig.~\ref{fig.9}.
}
\label{fig.8}
\end{figure}

As already shown in Fig.~\ref{fig.3b}(a), on short time scales one has
the typical dynamics of the single Cahn-Hilliard equation, i.e. a coarsening process.
However, the subsystem with larger wavelength coarsens faster than the other variable.
This process continues until the two variables $u$ and $v$  
lock into periodic patterns of the same characteristic length $L_c$ (see Fig.~\ref{fig.8}(a)).
Once the two variables have taken on the same wavelength, the coarsening process stops and $L_c$
remains constant. 
In this case the profile of the two variables are perfectly 
periodic with maxima and minima of both variables  occurring  in phase
as shown in Fig.~\ref{fig.9}(b). For comparison, the
evolution of the two uncoupled systems is reported in Fig.~\ref{fig.8}(b), where coarsening is not arrested during the simulation time window and the scaling of $L_c$
grows logarithmically in time as expected for a one-dimensional
CH equations in the absence of noise~\cite{pol_torc}. 
Furthermore, the spatial configurations for the variables $u$ and $v$ do not lock even at 
very long times, see Fig.~\ref{fig.8}(b).

\begin{figure}
\includegraphics[width=3.in]{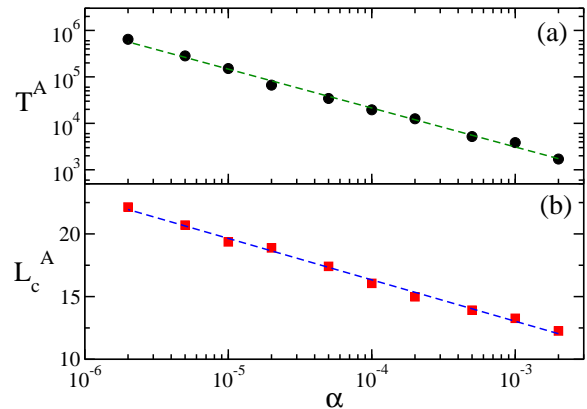}
\caption{Average arrest time $T^A$ (a) and average arrest length $L^A_c$ (b) as
a function of the coupling $\alpha$. The averages have been performed over
$20-60$ different initial conditions. The standard deviation measured in the
case of $T^A$ are of the order of the averages and in the case of $L^A_c$
of the order of $10 \%$ of the averages.  The dashed lines indicate
power-law (logarithmic) fitting to the data for $T^A$ ($L^A_c$),
namely $T_A \simeq 9.5\alpha^{-0.84}$ ($L_A^c \simeq  3.123 - 1.435 \ln(\alpha)$).
The system size and integration details are as in Fig.~\ref{fig.9}.
}
\label{fig.10}
\end{figure}

We have also investigated the scaling of the time of arrest $T^A$ of coarsening
with the coupling parameter $\alpha$. As shown in Fig.~\ref{fig.10}(a), a power-law scaling of the type 
$T^A  \simeq \alpha^{-\eta}$, with $\eta \simeq 0.8-0.9$, is observable.
Furthermore, the coarsening process is arrested at increasing characteristic
lengths $L^A_c$  diverging as $\ln(1/\alpha)$ for decreasing $\alpha$-values,
see Fig.~\ref{fig.10}(b). These two scaling laws are essentially
consistent with the logarithmic coarsening in time reported for the deterministic
one dimensional CH equation, thus suggesting that the asymptotic
value for the exponent $\eta$ should be one.

An arrest of coarsening has been previously reported for scalar fields in one
spatial dimension for the Oono-Shiwa model~\cite{Oono}. This is 
a modified version of the single CH equation, with an additional linear coupling 
to the order parameter, that has been developed to mimic phase separation in block 
copolymer melts~\cite{Oono,Goldenfeld,Villain}.
The analysis of this model
revealed that the system arrests and give rises to periodic regular stable
structures, similarly to what we observe here for the coupled CH equations.
However, while the arrest of coarsening in the
Oono and Shiwa model is due to the stabilization of long-wavelength modes by the additional linear term in the Oono-Shiwa model, 
this effect is not present for the coupled CH equations studied here.  
Instead, the arrest of coarsening stems from the interaction of the $u$ and $v$ 
fields. The presence of the initially longer wavelength mode in $u$ accelerates 
the coarsening  of initially the shorter wavelength modes in  $v$ much more than 
vice versa. Hence, one can say that each pattern acts as a template for the 
other one and as a result coarsening stops.  This is reminiscent of domain 
pinning seen in dewetting processes on heterogeneous substrates 
\cite{brusch2002,thiele2003}.

\section{Conclusions}

We examined coupled Swift-Hohenberg (SH) and the Cahn-Hilliard (CH) equations as 
simple models for pattern formation in systems with competing instabilities of 
different characteristic wavelength.
While a single SH o CH equation exhibits only instabilities connected with real 
eigenvalues and modes and, therefore, can only produce stationary periodic 
spatial patterns (SH) or a slow domain coarsening (CH), the asymmetric
coupling of two equations typically leads to the occurrence of oscillatory 
unstable modes with complex eigenvalues and, consequently, to the emergence of 
traveling waves and domains.

Using linear stability analysis we determine the conditions
for the occurrence of stationary and oscillatory instabilities and its dependence on the parameters
in the coupled SH and CH equations. In particular, the choice of opposite signs 
in the coupling produces the appearance of complex
eigenvalues  in the dispersion relations, see Eqs. (\ref{eq.SH.2.dr}) and (\ref{eq.CH.2.dr}).
At the same time this choice does not allow to rewrite the coupled system as a variational derivative 
of an associated energy functional.

The linear stability analysis is complemented by numerical simulations of the full nonlinear models
 in a wide parameter range.
These studies reveal a region of coexistence of traveling waves and Turing
patterns in the coupled SH equations. Furthermore, a rich variety of patterns has been
observed ranging from traveling waves with entrapped Turing patterns to one-dimensional 
defect turbulence.

On the other hand, for coupled CH equations the logarithmic coarsening,
typical of the one-dimensional single CH equations, is arrested even in presence
of a very small coupling. 
The spatial patterns  of the two variables  coarsen at different velocities. The pattern with smaller characteristic wavelength
coarsens faster than the one with longer wavelength until the profiles of the two patterns
attain the same spatial periodicity and are locked in space. 
As a result, coarsening stops and the wavenumber stays constant for arbitrary long times
The final domain pattern is stable and stationary in time. 
The corresponding final state for the two variables have the same
wavelength, but different amplitude profiles. Furthermore,
the asymptotic characteristic length diverges logarithmically for 
vanishingly small coupling.

Arrest of coarsening has been previously reported for scalar fields in one
spatial dimension for a modified version of the single specie CH equation
developed to mimic phase separation in block copolymer melts~\cite{Oono,Goldenfeld,Villain}.
In this case the final state reveals a periodic stable structure.
On the contrary, the arrest of coarsening reported in ~\cite{coarsening_arrest}, 
was associated to an unstable asymptotic pattern with a diverging amplitude.
In two dimensions arrest of coarsening has been shown in~\cite{glozer1995} for
spinodal decomposition of mixtures in presence of
an externally controlled chemical reaction, in~\cite{cencini}
for a CH, where the order parameter is subjected to an external stirring, both for active and
passive mixtures, and in ~\cite{zimmermann} for a modified CH, where the order parameter is coupled
linearly  to the Langevin equations describing the dynamics of Janus particles.

The results obtained here are analogous to what is found in more complex models 
describing lipid and protein dynamics  at membranes of biological 
cells \cite{john2005b,meacci2005} where waves have been observed following 
similar mechanisms. Since the study of the single CH equation linearly coupled to the order parameter
has been motivated by phase separation in block copolymers melt~\cite{Oono}, 
the analysis of two linearly coupled CH equations can eventually find application in
the study of self-directed self-assembly observed in mixtures of copolymers and nanoparticles~\cite{nat2005}.
Possible extensions of this study may consider the 
conserved SH equation \cite{thiele2013} where higher spatial derivatives are 
included.

In summary, a simple linear asymmetric coupling between two nonlinear equations 
may produce substantial changes in the dynamics, producing waves, chaotic dynamics, 
hysteresis or arrest of  coarsening.

%acknowledgments
\acknowledgments
We thank S. Lepri and P. Politi  for an useful exchange of ideas
and M. Cencini for a careful reading of the manuscript
prior to the submission. We acknowledge financial support from 
the German Science Foundation DFG within the framework of SFB 910 "Control of self-organizing nonlinear systems``.
AT has been partially supported by the Italian MIUR project 
CRISIS LAB PNR 2011-2013.

\end{document}